\documentclass[aps,pra,preprint,showpacs]{revtex4} 
\usepackage{amsmath,amssymb}  
\usepackage{rotating,epsfig}   
\usepackage{graphicx,fancybox}
\usepackage{color,calc,array}

\newcommand{\Beta}{\boldsymbol{\eta}}
\newcommand{\Brho}{\boldsymbol{\rho}}
\newcommand{\Bk}{\mathbf{k}}
\newcommand{\Bq}{\mathbf{q}} 
\newcommand{\BR}{\mathbf{R}}
\newcommand{\Bv}{\mathbf{v}}

\newcommand{\drho}{\mathrm{d}\Brho\;}
\begin{document} 

\title{Differential cross sections for single ionization of Li in collisions with fast protons and O$^{8+}$ ions}

\author{L. Guly\'as}
\affiliation{Institute of Nuclear Research of the Hungarian Academy
         of Sciences (ATOMKI), H-4001 Debrecen, P O Box 51, Hungary}
\author{S. Egri}
\affiliation{Department of Experimental Physics, University of Debrecen, 18/a Bem t\'er, H-4026 Debrecen, Hungary}         
\author{T. Kirchner}
\affiliation{Department of Physics and Astronomy, York University, 4700 Keele Street, Toronto, Ontario, Canada M3J 1P3}

%\submitto{\jpb}

\begin{abstract}
We study the process of single ionization of Li in collisions with H$^+$ and  O$^{8+}$ projectile ions at 6 MeV and 1.5-MeV/amu impact energies, respectively. Using the frameworks of the independent-electron model and the impact parameter picture, fully (FDCS) and doubly (DDCS) differential cross sections are evaluated in the continuum distorted-wave with eikonal initial-state approximation. Comparisons are made with the recent measurements of LaForge \textit{et al} [J. Phys. B \textbf{46} 031001 (2013)] for the DDCS and Hubele \textit{et al} [Phys. Rev. Lett. \textbf{110} 133201 (2013)] for the FDCS, respectively. For  O$^{8+}$ impact inclusion of the heavy particle (NN) interaction in the calculations is crucial and effects of polarization due to the presence of the projectile ion have also to be taken into account for getting very good agreement with the measured data. Our calculation reproduces the satellite peak structure seen in the FDCS for the Li(2s) measurement, which we explain as being formed by a combination of the binary and NN interactions.
\end{abstract}

\maketitle

\section{Introduction}
The study of single and multiple ionization of simple atoms by fast bare ion impact provides an excellent opportunity to explore mechanisms leading to the break up of few-body Coulomb systems \citep{sto97,M97,COLTRIM}.     
In the last few decades, thanks to the development of  cold-target recoil-ion momentum spectroscopy (COLTRIMS) \citep{U03}, there have been very intense efforts to explore the different mechanisms in fine details, see \cite{schulz05,schulz09,Shoffler09} and references therein. The very recent implementation of laser cooling in a magneto-optical trap combined with a reaction  microscope (MOTReMi) has opened up the possibility of studying collision processes with state-prepared target atoms \citep{Fischer2012}. 

Studying simple collision systems has the advantage of being able to get  complete kinematic information on the processes experimentally. Fully differential cross sections can be determined, whose interpretation offer a real challenge for theoretical modelling. In recent years, intense discussions have been generated e.g. on the role of the nucleus-nucleus (NN) interaction or on projectile coherence effects  which remain hidden in most of the less differential measurements \citep{Schulz07,Egodapitiya11,Sharma14}. The decisive role of the NN interaction has also been demonstrated in a recent kinematically complete experiment for single ionization in an initial-state selective study of   O$^{8+}$-Li(2s),Li(2p) collisions \citep{Hubele2013,LaForge2013}. Significant initial state dependence has been reported for the doubly differential cross section as function of electron energy and transverse momentum transfer. The experimental data were confronted with predictions from continuum distorted wave with eikonal initial state (CDW-EIS) calculations and, surprisingly, a classical description of the NN interaction provided the best agreement. In subsequent theoretical studies based on  the close-coupling approach (CC) \citep{Ciappina13} and the coupled-pseudostate approximation (CP) \citep{WW2014} noticeable effects of the NN interaction have been confirmed and reasonable agreement between experiment and theory was concluded.  

Distortion or polarization of an atomic electron orbital by other target electron(s) or by the incident charged particle is one of the most difficult task to deal with in the theoretical descriptions. The problem can be addressed in the distorted wave formalism \citep{crorev92} or in term of a polarization potential \citep{Joachain75}. To describe the polarization potential, several functional forms have been suggested  \citep{Nakanishi86,Zhang1992,Mitroy2010}. All of them have the  $V_{pol}\approx \alpha / 2r^4$ asymptotic behaviour at large distances $r$ from the target, where $\alpha$ is the dipole polarizability  constant. However, at shorter distances, the exact from of the potential is not available, which is the main reason for the existence of several analytical expressions. They are usually obtained by fits to experimental data or by using some reasonable assumption on the behaviour of the potential at short distances. 
A large number of theoretical studies have been devoted to selecting and testing  polarization potentials in the area of electron-atom scattering, see e.g. \citep{Nakanishi86,Zhang1992,Mitroy2010,Yurova2014} and references therein.
However, only little is known about the effects of $V_{pol}$ in heavy particle collisions, especially at high projectile energies. 

In this contribution we apply the  CDW-EIS method \citep{GFS95,gul08} to calculate the differential cross sections for single ionization of Li in collisions with 6 MeV  H$^+$ and 1.5 MeV/amu O$^{8+}$ projectile ions and discuss results in comparison with experimental data of LaForge \textit{et al} \citep{LaForge2013} and Hubele \textit{et al.} \citep{Hubele2013} and with available theoretical results. Using the independent electron picture the one-electron transition amplitudes are determined in the CDW-EIS model. Having the impact parameter dependent transition amplitudes the effects of the NN interaction are taken into account by a phase factor. No exact form of this factor is available, however, different assumptions imposed on it proved to be useful for exploring the mechanisms of the underlining processes. Special attention is paid to the role of $V_{pol}$. Different analytical forms are applied and a characteristic role of $V_{pol}$ is found at large momentum transfer values.  

The article is organized as follows. In Sec \ref{sec:theory} we summarize the main points of our theoretical description. In Sec \ref{sec:res.} the results are discussed. A summarizing discussion is provided in Sec. \ref{sec:conc.}.
Atomic units characterized by $h = m_e = e = 4\pi \epsilon_0 = 1$ are
used unless otherwise stated.

\section{Theory}
\label{sec:theory}

In order to reduce the very challenging many-electron treatment to a much simpler one-electron description we consider only one electron in Li as active over the course of the collision while the others remain bound to  their initial state. The non-active electrons are taken into account by an effective potential $V_{\mathrm{Li}}$ that represents the interactions in the (1s$^2$2s) ground state configuration. This potential is obtained from the exchange-only version of the optimized potential method (OPM) of density functional theory, i.e. it includes the electron nucleus Coulomb interaction, screening, and exchange terms exactly and exhibits the correct -1/r$_T$ behaviour, but it neglects electron correlation ($\mathrm{r}_T$ denotes the position vector of the electron relative to the target nucleus) \cite{Engel-opm}. The above assumption on the description of the  target is the essential point in the application of the independent electron model (IEM) where electrons are considered to evolve independently and it enables to simplify the treatment of a many-electron  collision problem to a three-body system \cite{M97}.         

In the following we consider a three body-collision, where a bare projectile ionises a target initially consisting of a bound system of an electron and a core represented by the $V_{\mathrm{Li}}$ interaction potential. 
Furthermore, we apply the impact parameter method, where the projectile follows a
straight line trajectory $\BR=\Brho + \Bv t$, characterized by the constant
velocity $\Bv$ and the impact parameter $\Brho \equiv (\rho,\varphi_\rho)$ \cite{mcd70}.
The one electron Hamiltonian has the form
\begin{equation}
%\everymath{\displaystyle}
h(t)= -\frac{1}{2} \Delta_{\mathbf{r}_T} + V_{\mathrm{Li}}(|\mathbf{r}_T|) - \frac{Z_P}{\mathbf{r}_P},
\label{el-ham}
\end{equation}
where $\mathbf{r}_P$ denotes the position vector of the electron relative to the projectile nucleus having nuclear charge $Z_P$. 
The single  particle scattering equation is solved within the framework of the CDW-EIS approximation, where unperturbed atomic orbitals in both the incoming and outgoing channels have been evaluated on the same $V_{\mathrm{Li}}$ potential, see refs. \cite{crorev92,GFS95,fain96,gul08} for more details. Here we note that a similar formalism, including transition amplitudes from a basis generator method calculation, was used in our recent study of excitation and ionization in the 1.5-MeV/amu O$^{8+}$ - Li collision system \citep{gul14}. 
 
The doubly differential cross section (DDCS) differential in energy of the emitted electron $E_e$ (=$k_e^2/2$; $\Bk_e \equiv (k_e,\theta_e, \varphi_e)$ is the electron momentum) and in the transverse component ($\Beta \equiv (\eta, \varphi_{\eta})$) of the projectile's momentum transfer $\Bq = \Bk_i - \Bk_f = -\Beta + \Delta E /v$ is given as
\begin{equation}
%\everymath{\displaystyle}
 \frac{ \mathrm{d} \sigma^{2}}{ \mathrm{d} E_e \mathrm{d}\eta} = k_e \eta
\int_{-1}^{+1} \mathrm{d} (\cos \theta_e)  \int_{0}^{2\pi} \mathrm{d} \varphi_e
\int_{0}^{2\pi} \mathrm{d} \varphi_{\eta}
|\mathcal{R}_{i \Bk}(\Beta)|^2,
\label{ddcs}
\end{equation}
where $\Delta E= E_e-\varepsilon_i$, $\varepsilon_i$ is the binding energy of the electron in the initial state, $\Bk_{i (f)}$ stands for the projectile momentum before (after) the collision and $\mathcal{R}_{i \Bk}(\Beta)$ is the transition matrix.

In equation (\ref{ddcs}) the projectile's momentum transfer and consequently the projectile scattering is defined by the interaction of the projectile with the active electron. However, the scattering of the projectile also depends on its interaction with the target core, (so-called NN interaction). We approximate this interaction by using the potential
\begin{equation}
%\everymath{\displaystyle}
V_{\mathrm{NN}}(R)=Z_P Z_T/R + V_s(R) + V_{pol}(R),  
\label{VZZ}
\end{equation}
where 
\begin{equation}
%\everymath{\displaystyle}
V_s(R)= Z_P \sum_{i=1}^{2}\left< \psi^i_{1s} |- 1/|{\bf R} - {\bf r_i}| |\psi^i_{1s} \right>. 
\label{scren}
\end{equation}
describes the interaction between the projectile and the passive electrons. In (\ref{scren}) $\psi^i_{1s}$ is approximated by a hydrogenlike wave function ($\psi^i_{1s}=N_ie^{-z_e^i r_T}$) 
with effective charge $z^i_e(=2.65)$ (Slaters's rule, \cite{Slater30}). Taking  $z^1_e=z^2_e$, we obtain
\begin{equation}
%\everymath{\displaystyle}
 V_s(R) = - 2 Z_P \left[ 1/R - ((1+z_e R)/R)e^{-2z_eR} \right].
 \label{screnz}
\end{equation}
On the accuracy of (\ref{screnz}) we note that we have evaluated $V_s(R)$  with the Li(1s)  OPM orbital and have compared it with (\ref{screnz}). The difference is very small, which is supported by the fact that the OPM  Hartree potential  $V_{H}^{OPM}$ for the Li$^+$(1s$^2$) configuration has the limit: $\lim_{R \to 0} V_H^{OPM}(R) \to$ 5.375 which corresponds to $z_e$=2.687.
It can be checked that $\lim_{R \to 0} V_{NN}(R) \to 3 Z_P/R$ and $\lim_{R \to \infty} V_{NN}(R) \to Z_P/R$ for $Z_T$=3.

$V_{pol}(R)$ in (\ref{VZZ}) accounts for the (adiabatic) polarization or distortion of the core electrons by the incident charged particle \cite{Joachain75,Yurova2014}. 
{\color{black}
Its use is based on the idea that the electric field of the projectile at distance R gives rise to an instantaneous
(first-order) distortion of the core-electron orbitals, thereby modifying the interaction of those electrons with 
the projectile. Polarization potentials have been used in many studies up to fairly high projectile energies \cite{Mitroy2010,Yurova2014,Nakanishi86,Schrader79};
in particular in \citep{WW2014}, in which the CP method was applied to the collision system
considered in this work.
% The present derivation of $V_{pol}$ rests on static electric filed which applied commonly in describing collision processes from  low to high (not relativistic) impact energies \cite{Yurova2014,WW2014,Nakanishi86,Schrader79}.  The use of frequency-dependent polarizability might be reasonable for much higher impact velocity and with highly charged projectiles \citep{Mitroy2010,Voitkiv2001,Voitkiv03}
}
Different types of analytical approximations are available for $V_{pol}(R)$ and we consider the following  frequently used three forms:  
\begin{equation}
%\everymath{\displaystyle}
V_{pol}(R)=-\frac{\alpha Z_P^2}{2(R^2+d^2)^2},  
\label{VP}
\end{equation}  
where $\alpha$ is the atomic dipole polarizability parameter \cite{Mitroy2010} and $d$ is a "cut-off" parameter whose value is taken as the  average radius of the Li$^+$(1s$^2$) ion (d=0.57 a.u. for the present case) \cite{Zhang1992};
\begin{equation}
%\everymath{\displaystyle}
V_{pol}(R)=-\frac{\alpha Z_P^2}{2R^4}(1-\exp (-(R/r_d)^6) ),  
\label{VPExp}
\end{equation}
with $r_d=$0.47 \cite{Bottcher1971};
and
\begin{equation}
%\everymath{\displaystyle}
V_{pol}(R)=-\frac{\alpha Z_P^2}{2R^4}(1-e_n({\xi})\exp(\xi))^m,  
\label{VPExp-e}
\end{equation}
with $\xi=R/r_o$ where $e_n({\xi})$ is the truncated exponential function $e_n=\sum_{i=0}^{n}(\xi)^i/i!$, $r_o$=0.116, $m=$6 and $n$=2  \cite{Nakanishi86}.
All these polarization potentials have the form $V_{pol}(R)\approx - \alpha /2R^4$ at large distances and differ in the short-range limit due to the "cut-off" parameters or functions which contain parameters estimated on some reasonable assumptions. 
  
Effects of the NN  interaction on the scattering process can be investigated by solving the Schr\"odinger equation for the Hamiltonian (\ref{el-ham}) with inclusion of the potential (\ref{VZZ}). However, the solution simplifies remarkably if one considers that (\ref{VZZ}) depends on $R$ alone and so $V_{NN}$ can be removed from (\ref{el-ham}) by a phase transformation. The transition matrix $\mathcal{R}_{i \Bk}(\Beta)$ that takes the internuclear interaction into account can then be expressed as \cite{mcd70} 
\begin{equation}
%\everymath{\displaystyle}
\begin{array}{c}
 \mathcal{R}_{i \Bk}(\Beta) =\frac{1}{2 \pi} \int \drho e^{i \Beta \cdot \Brho} 
   a_{i \Bk}(\Brho) 
\label{rifn}
\end{array} 
\end{equation} 
with $ a_{i \Bk}(\Brho)= e^{i \delta(\rho)} \mathcal{A}_{i \Bk}(\Brho)$, where
$\mathcal{A}_{i \Bk}(\Brho)$ is the transition amplitude calculated without the internuclear interaction, and the phase due to (\ref{VZZ}) is expressed as
\begin{equation}
%\everymath{\displaystyle}
 \delta(\rho) =- \int_{-\infty}^{+\infty} \mathrm{d} t V_{NN}(R(t)). 
\label{nnphase}
\end{equation}

\section{Results}
\label{sec:res.}

\subsection{Doubly differential cross sections for the ionization of Li (2s) and Li(2p)}
 
In Figures (\ref{fig1})-(\ref{fig6}) we compare our CDW-EIS results of 
$\mathrm{d} \sigma^{2} / \mathrm{d} E_e \mathrm{d}\eta$ for proton  and $O^{8+}$ impact on Li(2s) and Li(2p) with measurements of LaForge  $\textit{et al}$ \cite{LaForge2013} and Hubele \textit{et al} \citep{Hubele2013}  for the electron ejection energies of 2, 10 and 20 eV. The experimental data of \cite{LaForge2013} and \citep{Hubele2013} are  not on the absolute scale, only the relative normalisation was fixed for different  $E_e$ and for Li(2p) relative to Li(2s). Following previous works \citep{LaForge2013,Ciappina13,WW2014} we have fixed the absolute scale in the Figures by normalising the data to our CDW-EIS cross sections for 2eV ejection from Li(2s) at $\eta$=0.65 a.u.

\subsubsection{H$^+$ projectile impact at 6 MeV }

In is clear from Figs (\ref{fig1}) -(\ref{fig3}) (a) and (b) that the role of the internuclear interaction is negligible in the whole range of $\eta$ where the measurements have been taken when the projectile is a 6 MeV proton. This has also been noted in \cite{LaForge2013,WW2014}. We have also evaluated cross sections with $V_{NN}=Z_P Z_{eff}/R$, and  $Z_{eff}$=3.0; 1.0; and 1.35. The first and second choice correspond to a close and a distant collision of the heavy particles respectively and the last one is an intermediate situation in which  $Z_{eff}$ is obtained from  Slater's screening rule. DDCS's evaluated  with these NN interactions are not presented in the Figures (\ref{fig1})-(\ref{fig3}) (a) and (b) as they are almost the same as those obtained with or without (\ref{VZZ}). The indifference of the DDCS to the form of NN interaction is explained by the almost constant character of the internuclear  phase over the region of $\rho$ (see (\ref{nnphase})), where  $\mathcal{A}_{i \Bk}(\Brho)$ has significant values.   
%This form of nuclear-nuclear (NN) interaction is qu

Panels (a) and (b) of Figures (\ref{fig1}) -(\ref{fig3}) also show  results of CDW-EIS calculations by LaForge $\textit{et al}$ \cite{LaForge2013}. Their results are almost the same as the present ones for all  $E_e$'s when the shapes of the curves  are considered. Slight differences appear mostly at small $\eta$ values and for those $\eta$'s where the DDCS's have maxima. As noted above the NN interaction is negligible for proton impact, so the difference between the two CDW-EIS calculations lies in the description of the target orbitals. 
%The CDW-EIS calculations by LaForge $\textit{et al}$ \cite{LaForge2013} have been performed with screened H-like wave functions for Li. 
However, it must be noted that the above discrepancies are within the experimental uncertainties. 

\subsubsection{O$^{8+}$ projectile impact at 1.5 MeV/amu }

\begin{figure}[ht]
\centering
\epsfig{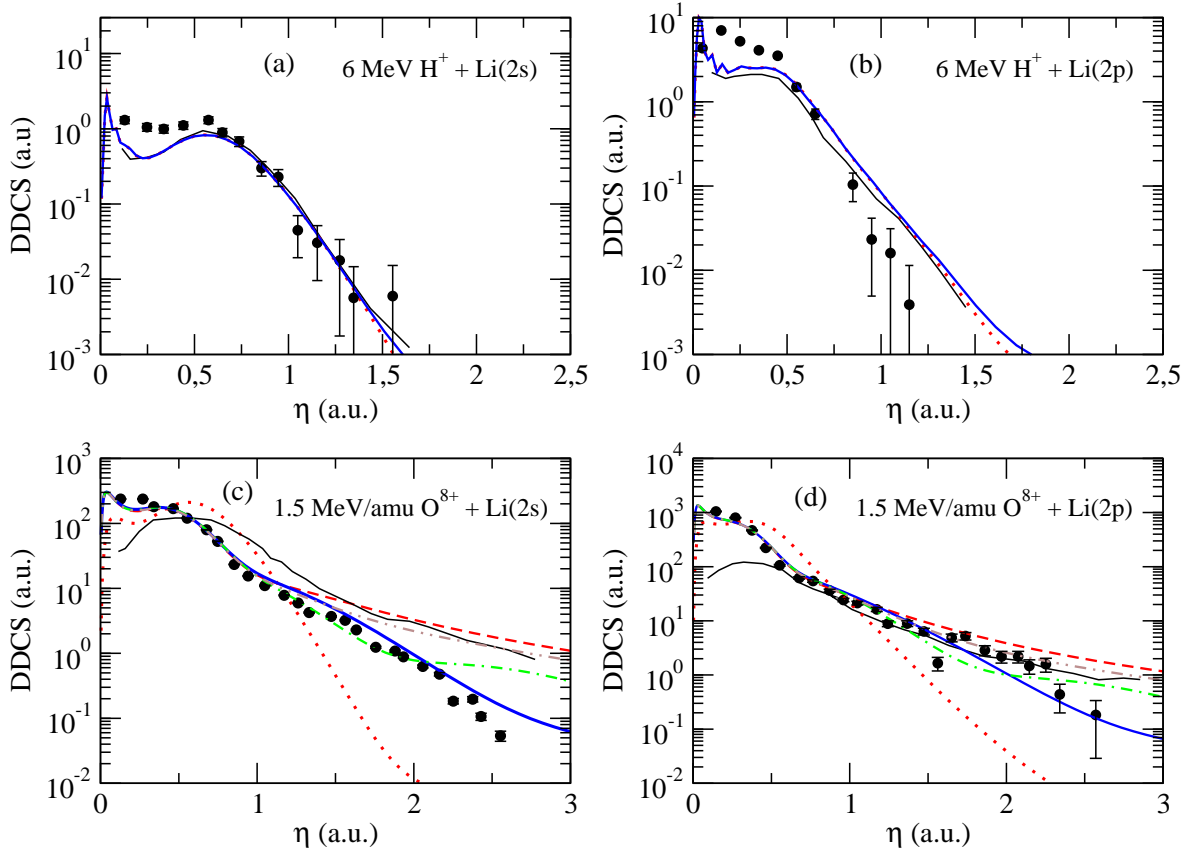}
\caption{
(Color online) DDCS for single ionization of Li(2s)  and Li(2p) by 6 MeV H$^+$ (upper panels)
and 1.5 MeV/amu O$^{8+}$ (lower panels) impact as a function of $\eta$ for E$_{el}$=2 eV.
Theories: present results, dotted red lines represent calculations without the NN interaction; calculations with NN interaction where V$_p$=0 are dashed red lines and where V$_p$ is given by (\ref{VP}) with $d$=0.57 are thick solid blue lines; (\ref{VPExp}) are dot-dashed green lines; (\ref{VPExp-e}) are dot-dot-dashed brown lines, see the text. Black thin solid curves are CDW-EIS calculations by LaForge \textit{et al} with classical NN interaction \cite{LaForge2013}. Experiment: $\bullet$ from \cite{LaForge2013} as renormalized recently \citep{Fischer-14}.  
}
\label{fig1}
\end{figure}

\begin{figure}[ht]
\centering
\epsfig{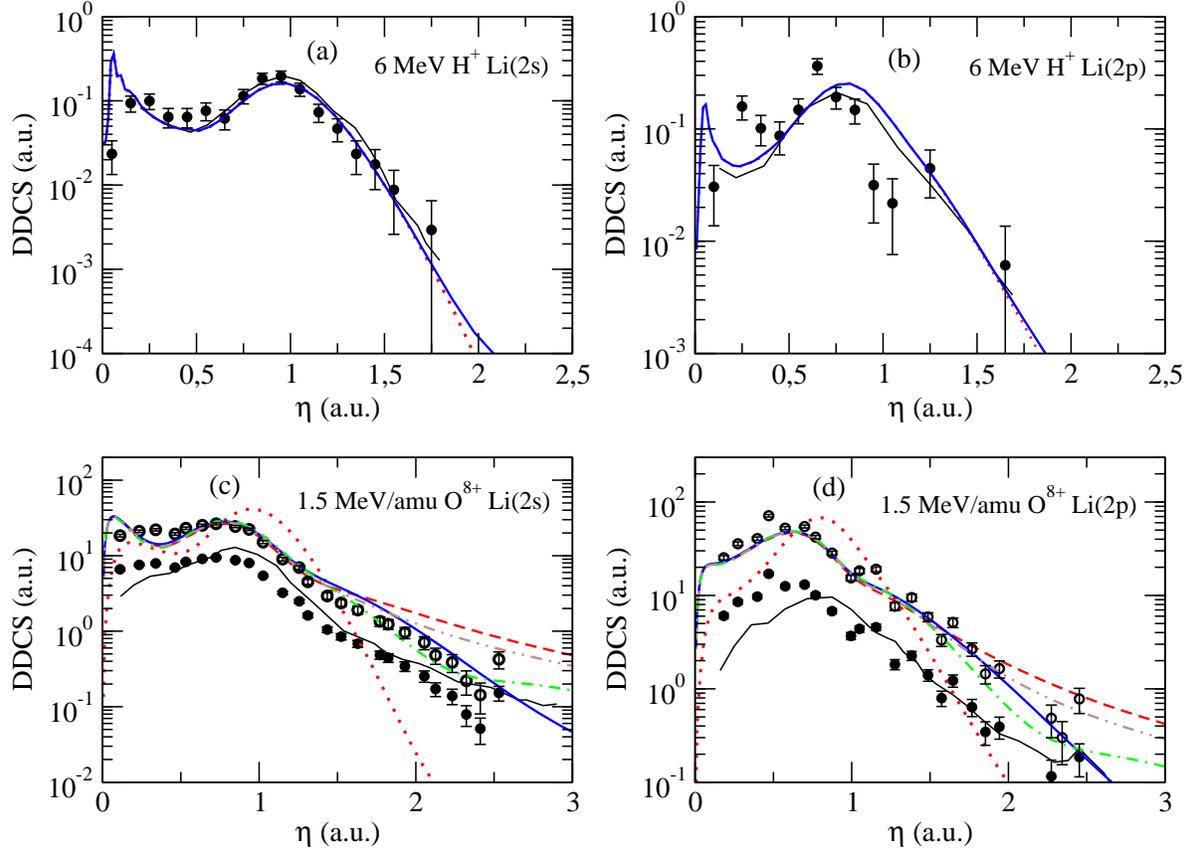}
\caption{
(Color online) Same as Figure (\ref{fig1}) but for E$_{el}$=10 eV. $\circ$ are the multiplied experimental data for the best visual fit with the theory.   
}
\label{fig2}
\end{figure}

\begin{figure}[ht]
\centering
\epsfig{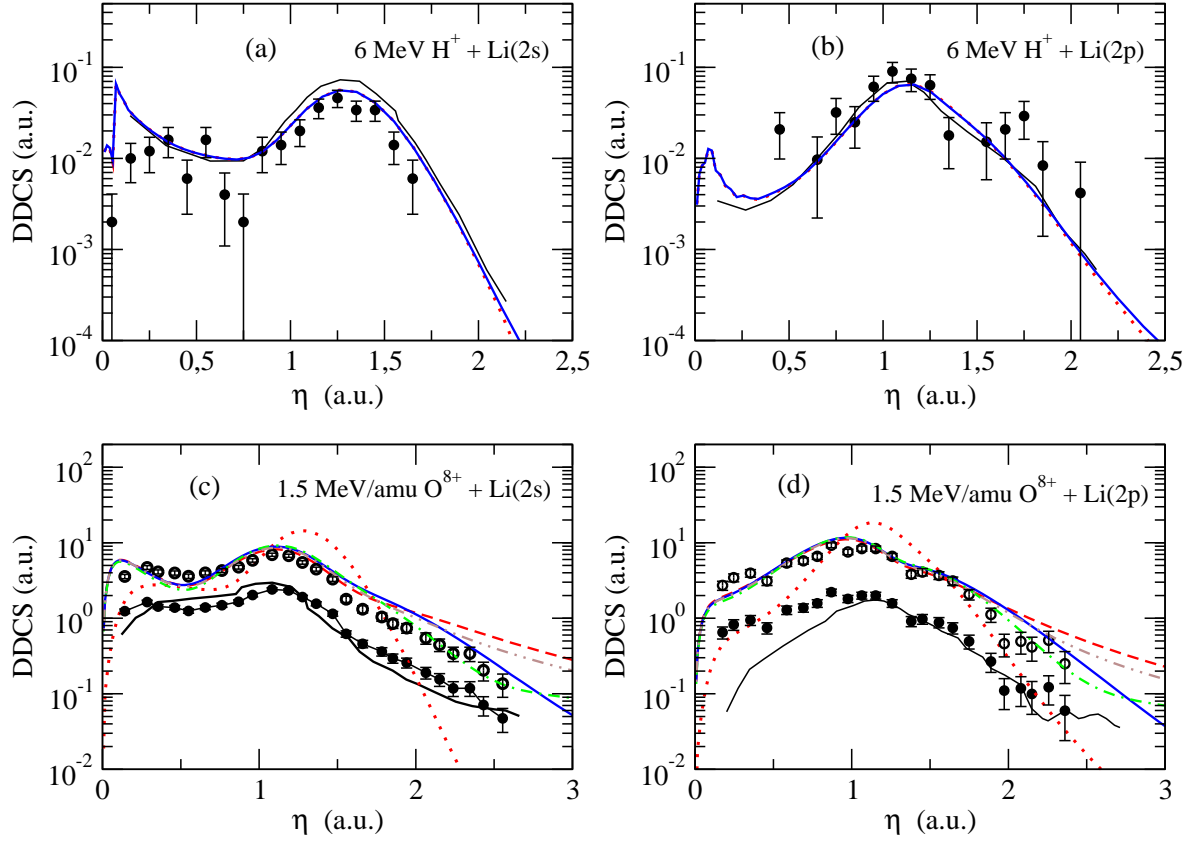}
\caption{
(Color online) Same as  Figure (\ref{fig2}) but for E$_{el}$=20 eV.
}
\label{fig3}
\end{figure}

The picture becomes more complicated for O$^{8+}$ ion impact, see the lower panels ((c) and (d)) in Figures (\ref{fig1}) -(\ref{fig3}). The role of the internuclear interaction is more evident for this projectile than for the H$^+$ ion. DDCS's evaluated with and without the NN interaction differ considerably. $\delta(\rho)$, see (\ref{nnphase}), oscillates rapidly with $\rho$ when  $Z_p$=8  and the DDCS is sensitive to the form of the NN interaction. Let us first consider results evaluated with the NN interaction where the polarization potential is set to zero, $V_{pol}$=0 in  (\ref{VZZ}). Compared to the measurements, calculations with the  more sophisticated  screening potential (\ref{scren}) present reasonable agreement. For all $E_e$ and for both 2s and 2p the agreement is very good from medium to low $\eta$ values and discrepancies appear only in the large $\eta$ region where the calculations overestimate the measurements. Calculations with $V_{NN}=Z_P Z_T/R$ (not shown in the Figures) fail in almost the entire range of $\eta$ showing the important screening role of the passive electrons. At the same time when the internuclear interaction is taken into account by $V_{NN}=Z_P/R$ the calculations result in DDCS which are very similar to those obtained with (\ref{VZZ}), for the case of $V_{pol}$=0. This becomes clear if we consider the range of $\rho$ over  which $ \mathcal{A}_{i \Bk}(\Brho)$ has significant values. This range extends to $\rho \approx$ 25-30 a.u., however,  $V_{NN}(R)$ of (\ref{VZZ})  reaches its asymptotic limit and behaves as $Z_P/R$  when $\rho \geq$ 1-2 a.u. So the DDCS is governed by the asymptotic form of the NN interaction in almost the entire range of impact parameters. {\color{black} Here we note that similar findings on the role $V_s$ in (\ref{VZZ}) were reported in \citep{Voitkiv09} for the case of 100 MeV/amu C$^{6+}$ and 1 GeV/amu  U$^{92+}$ impact on helium. }

Let us now consider results obtained with (\ref{VZZ}), i.e. the full form of  the NN interaction potential, in which $V_{pol}$ is also taken into account. It is clear from  the lower panels of Figures (\ref{fig1}) -(\ref{fig3}) that $V_{pol}$ has a negligible effect at low $\eta$ values, however, it plays a  drastic role in the large $\eta$ region. Very good results are found in the whole $\eta$ region for all electron energies when $V_{pol}$ takes the form (\ref{VP}). Taking (\ref{VPExp}) for $V_{pol}$ also results in good agreement below $\eta \leq$ 2 a.u., while above this value  these calculations overestimate the experiment. 
We note that a $V_{pol}$ same or similar as in Eq. (\ref{VPExp}) was probably used in \citep{WW2014}, which is supported by the fact that the results reported in that work are consistent with the present ones.
%We note that $V_{pol}$ with the same or similar form as of (\ref{VPExp}) %might also been used in \cite{WW2014}, which is supported by that the %results reported in that work  are consistent with the present ones.
Results with $V_{pol}$ of (\ref{VPExp-e}) are very close to those  obtained with $V_{pol}$=0. 

As for the case of H$^+$ impact, panels (c) and (d) of Figs. (\ref{fig1})-(\ref{fig3}) also present the  CDW-EIS results  of La Forge \textit{et. al.} \cite{LaForge2013}. Their CDW-EIS calculations, in which the NN interaction was taken into account classically, fail to reproduce the measured data mostly at low $\eta$ values. La Forge \textit{et. al.} \cite{LaForge2013} also presented results where the NN interaction was accounted for quantum mechanically in terms of the eikonal approximation.  However, they found that the classical treatment was more adequate especially at large $\eta$ with increasing E$_{e}$. Given the good results of the present calculations we see no reason to perform similar calculations with a classical inclusion of the NN interaction. 
%We  also note that due to the dense oscillatory character of (\ref{nnphase}) the DDCS is more sensitive to the treatment of $ \mathcal{A}_{i \Bk}(\Brho)$, which was not the case for H$^+$. On that score we think that part of the differences between the present and treatment of La Forge \textit{et. al.} \cite{LaForge2013} is related to the description of target orbitals.  

Hitherto, we have discussed DDCS results by considering only their shapes.
Obviously, confronting a calculation with a measurement that is lacking an absolute normalization might influence the assessment  of the validity of the theory. In Figures (\ref{fig1})-(\ref{fig3}), we have normalized the measured DDCS to our calculation at $\eta$=0.65 a.u. for Li(2s) at $E_e$=2 eV. Normalization at a different $\eta$ might modify the judgement of the theory. This is especially true when relative cross sections for the different ejection energies are considered. 
%A drift in the normalization of the DDCS with  increasing emission energy has also been noted in \cite{WW2014}. 
These comments mainly apply to the case of O$^{8+}$ impact, see Figs. (\ref{fig2}) and (\ref{fig3}), where shifted experimental data showing the "best visual fit" are also presented with open symbols. The shifts correspond to factors of about 2-5 depending on the collision parameters.
%for the case of 2s initial state are almost the same as the deviation between measured and calculated  single differential cross sections presented in Figure (\ref{fig4}). For 2p different shift have been applied for the best visuals fits. 
A similar drift in the relative normalisation of the DDCS for O$^{8+}$ impact has also been noted in \cite{WW2014}.

\subsubsection{Exploring the role of the polarization potential}

\begin{figure}[ht]
\centering
\epsfig{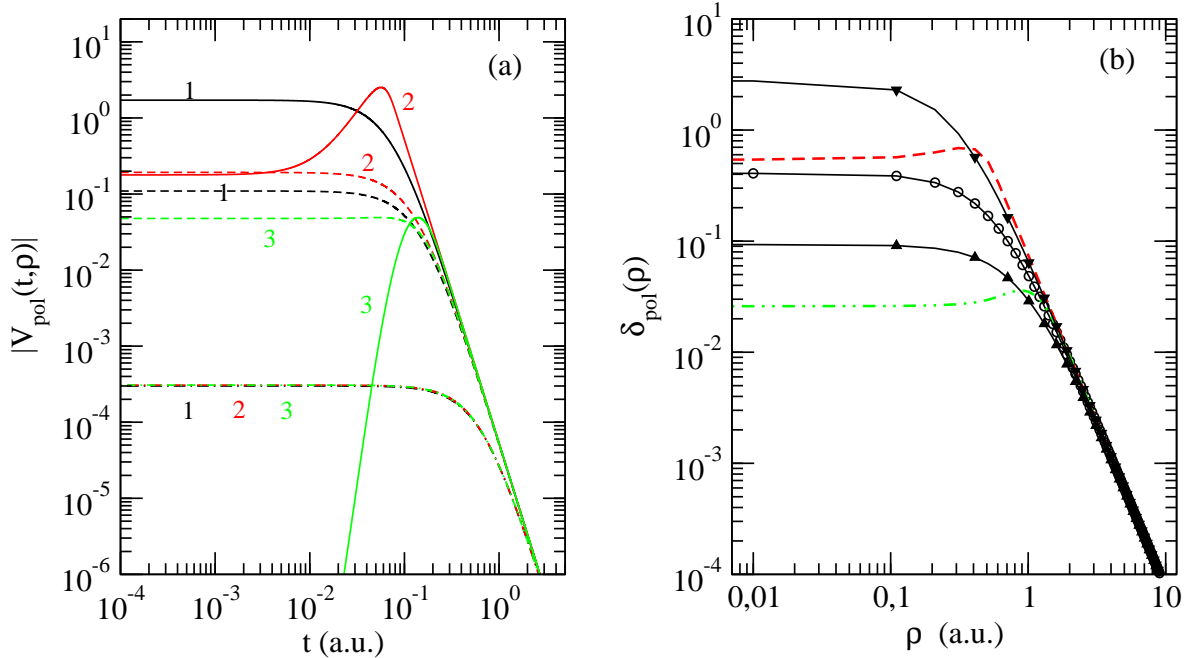}
\caption{
(Color online) (a) $|V_{pol}(R)|$ of (\ref{VP}) (d=0.57), (\ref{VPExp}) and (\ref{VPExp-e}) represented by curves 1, 2 and 3 as functions of $t=\frac{1}{v} \sqrt{R^2-\rho^2}$ for $\rho$=0.1, 1. and  5.0 a.u. with solid, dashed and dot-dashed lines respectively. (b) Internuclear phases (\ref{nnphase}) are shown due to $V_{pol}$ of (\ref{VP}) for $d$=0.57, 0.3 and 0.93 as solid lines with circle, triangle down and triangle up symbols; of (\ref{VPExp}) with dashed red and of (\ref{VPExp-e}) with dot-dot-dashed green lines.
}
\label{fig4}
\end{figure}

\begin{figure}[ht]
\centering
\epsfig{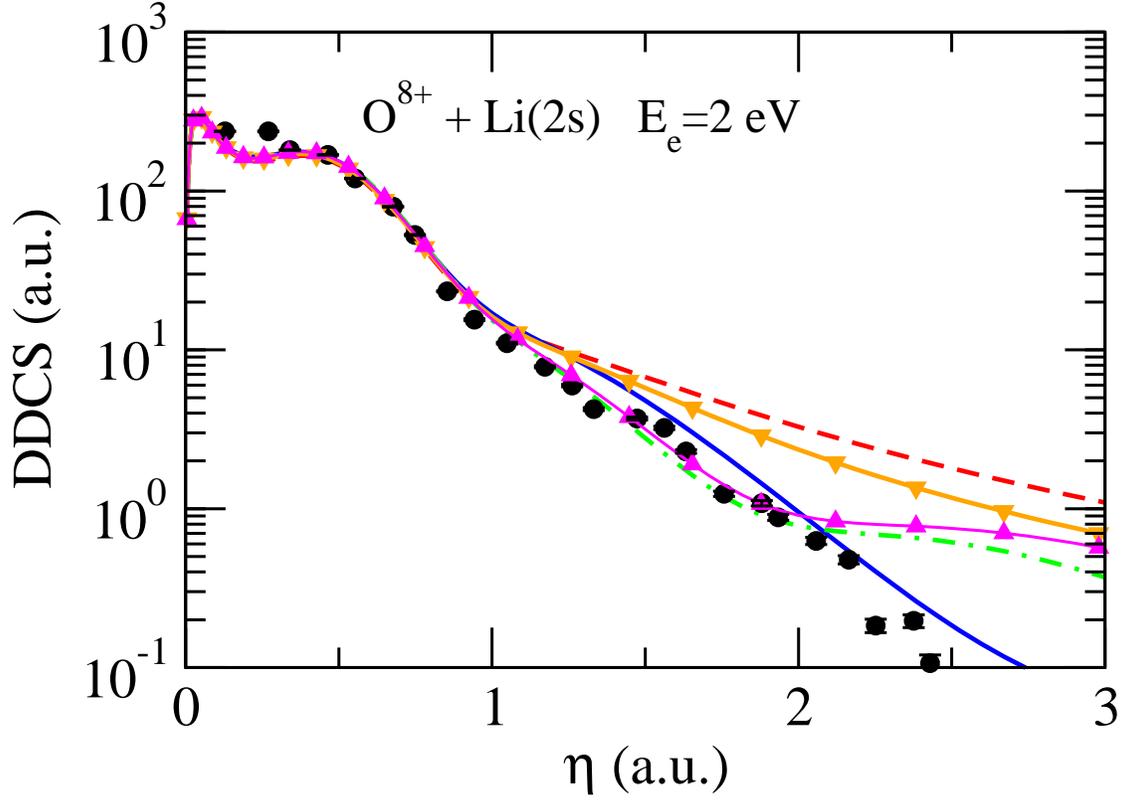}
\caption{
(Color online) DDCS for single ionization of Li(2s) by
1.5 MeV/amu O$^{8+}$ impact as a function of transverse momentum transfer for E$_{e}$=2 eV. Present calculations including the NN interaction with $V_{pol}$=0 are represented as dashed red and with $V_{pol}$ given by (\ref{VP}) with $d$=0.57 as solid blue,  d=0.3 as solid pink + triangle up, d=0.93 as solid orange + triangle down, and by  (\ref{VPExp}) as dot-dashed green lines, respectively.  $\bullet$ experimental data from \cite{LaForge2013} as renormalized recently \citep{Fischer-14}. 
}
\label{fig5}
\end{figure}

Deviations of DDCS's at large $\eta$ obtained with the different forms of $V_{pol}$ can be explored by considering the potentials and the internuclear phases ($\delta_{pol}$) evaluated only on them (see (\ref{nnphase})).  Figure (\ref{fig4}) (a) shows $V_{pol}$ of (\ref{VP})- (\ref{VPExp-e}) as a function of $t$ for three different $\rho$ values, note that $\rho$ and $t$ are related by $\BR=\Brho+\Bv t$. This Figure indicates that considerable differences among the potentials appear only for $\rho \leq$ 1 a.u.. $V_{pol}$ of (\ref{VP}) contains a "cut-off" parameter for which we take the 1s shell radius d=0.57 a.u.. Different criteria for d  have also been proposed in \citep{Yurova2014} and \citep{Zhang1992}. They result in d=0.3 and 0.93, respectively. $V_{pol}$ evaluated with these values of d are not represented in (\ref{fig4}) (a). However, deviations related to the use of different d parameters in (\ref{VP}) can be assessed from Figure (\ref{fig4}) (b). Figure (\ref{fig4}) (b) presents $\delta_{pol}$ evaluated with $V_{pol}$ of (\ref{VP}) using three different values for d in comparison with those obtained from $V_{pol}$ of (\ref{VPExp}) and (\ref{VPExp-e}). It is clear that the phases are the same for all forms of polarization potential when $\rho \geq$ 1, however, the deviations are significant at small $\rho$ due to different cutting procedures in (\ref{VP}) and  different forms of $V_{pol}$.
Apart from the very low $\rho$ region, which is unimportant for the DDCS, $V_{pol}$ of (\ref{VP}) with d=0.3  provides nearly the same $\delta_{pol}$ as that of (\ref{VPExp}). At the same time $\delta_{pol}$ provided by  $V_{pol}$ of (\ref{VP}) with d=0.93 and of (\ref{VPExp-e}) have very small values. This fact explains the small differences between DDCS's  with (\ref{VPExp-e}) and without polarization potential observed in Figures (\ref{fig1})-(\ref{fig3}) (c) and (d). It is also obvious that deviations in $\delta_{pol}$  and so in  $V_{pol}$ are manifested in the DDCS in the large $\eta$ region. This can be observed in Figure (\ref{fig5})  where the DDCS evaluated with different forms of $V_{pol}$ is presented for O$^{8+}$ + Li(2s) collsions for $E_e$=2 eV.

\begin{figure}[ht]
\centering
\epsfig{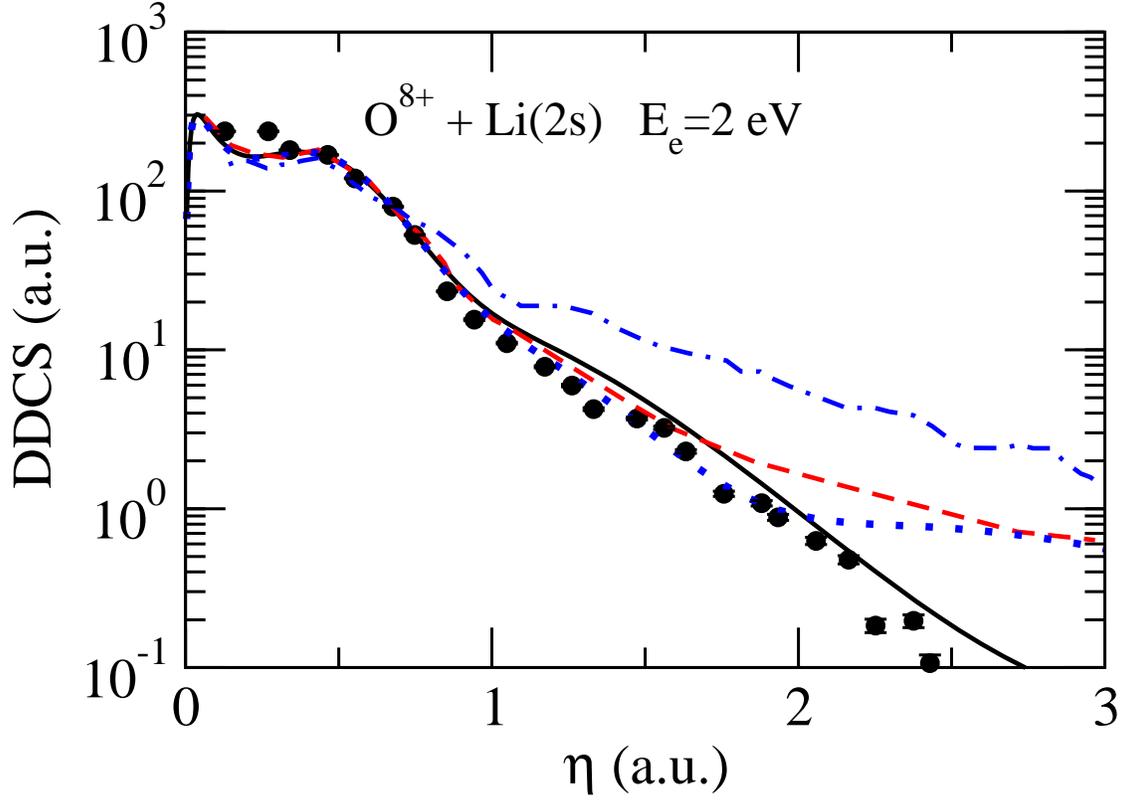}
\caption{
(Color online) DDCS  for single ionization of Li(2s) by
1.5 MeV/amu O$^{8+}$ impact as a function of transverse momentum transfer for E$_{e}$=2 eV.  Present calculations including the NN interaction with $V_{pol}$ given by (\ref{VP}) with $d$=0.57 and 0.3 are the solid black and dotted blue lines, respectively.  Dashed red and dot-dashed blue lines are the CP and CC calculations from \cite{WW2014} and \citep{Ciappina13}, (results from \citep{Ciappina13} are normalised to the present data at $\eta$=0.65 a.u.). $\bullet$ experimental data from \cite{LaForge2013}  as renormalized recently \citep{Fischer-14}. 
}
\label{fig5-a}
\end{figure}

\begin{figure}[ht]
\centering
\epsfig{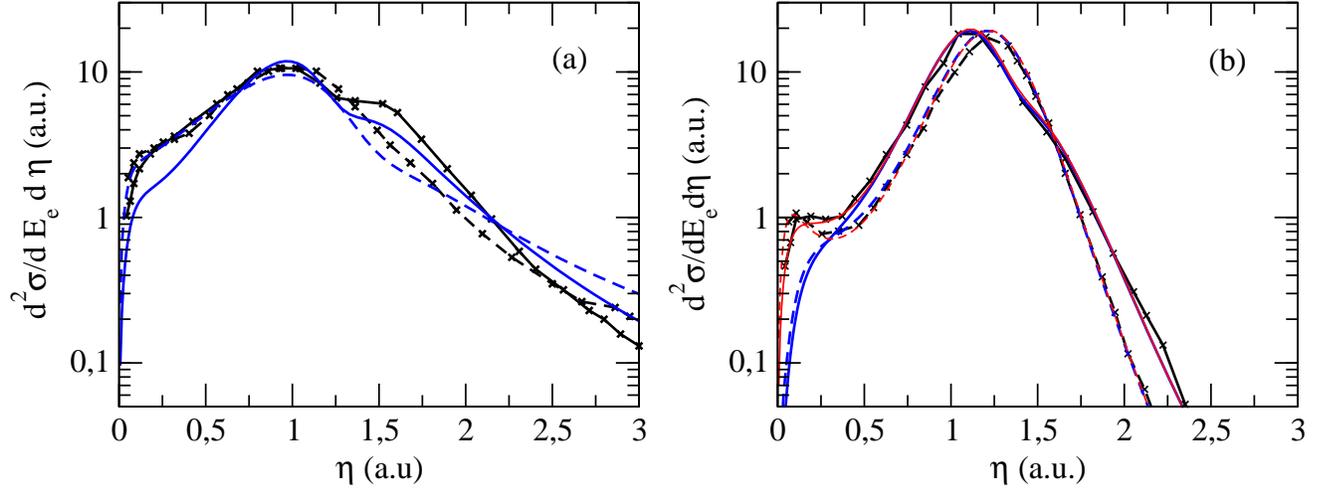}
\caption{
(Color online) DDCS for single ionization of  Li(2p$_o$) (dashed lines)  and Li(2p$_1$) (solid lines) by 1.5 MeV/amu O$^{8+}$ impact as a function of $\eta$ for E$_{e}$=20 eV with (a) and without (b) internuclear interaction. In (a), blue lines are present CDW-EIS results and black lines with crosses are CP results from \cite{WW2014}. In (b) blue thick and red thin lines are present CDW-EIS and B1 results, respectively, black lines with crosses are B1 results from \cite{WW2014}.   
}
\label{fig6}
\end{figure}

In \cite{WW2014} ionization of Li was discussed within the framework of the coupled pseudostate (CP) model. Their results show similar good agreement with the measurement as the present ones at low and medium $\eta$ values when the shapes of the DDCS's are considered. A similar study within the framework of the time-dependent close-coupling (TDCC) method was performed for $O^{8+}$ impact only in \citep{Ciappina13}. Reasonable agreement, especially for the Li(2p) target was observed. Results of these calculations are not included in Figures (\ref{fig1})-(\ref{fig3})) for the sake of clarity. However, in Figures (\ref{fig5-a}) and (\ref{fig6}) we give a comparison of the present, the CC, and the TDCC results for some selected collision parameters. In Figure (\ref{fig5-a}) we compare the present DDCS results for O$^{8+}$ + Li(2s) collisions at $E_e$=2 eV with the CP and the TDCC results. The DDCS from the TDCC calculation is normalized to the present data at $\eta$=0.65 a.u. while those from the CP is shown on an absolute scale.  In \cite{WW2014} the polarization potential was given by (\ref{VPExp}) and the agreement between their and our results obtained with $V_{pol}$ of (\ref{VPExp}) is good except around $\eta$=2 a.u.
% might be attributed to slight differences in the form of NN interactions.
In \citep{Ciappina13} the NN interaction was taken into account by the Coulomb repulsion with an effective charge and those results seem to be  comparable to our results without $V_{pol}$. 

Similarly good agreement between our and the CP results of \cite{WW2014} can be observed in Figure (\ref{fig6}) showing the DDCS for 1.5 MeV/amu O$^{8+}$ Li(2p$_{0,1}$) collisions at $E_e$=20 eV. It is seen that CP and CDW-EIS calculations agree well in the binary region and slight discrepancies appear at low and high $\eta$ values (see Figure (\ref{fig6}) (a)). It is important to note the good agreement on the absolute scale. Figure (\ref{fig6}) (b) shows results of calculations in which the NN interaction is neglected. Besides the present CDW-EIS results, results of a first Born (B1) calculation performed by us and a B1 calculation from \cite{WW2014} are also presented. The two B1 calculations are in very good agreement and differ from the CDW-EIS cross sections only at $\eta \le$ 0.5 a.u. values. This tells us that, except for the very low region of $\eta$, the collision with the O$^{8+}$ projectile is still in the B1 regime. Including the NN interaction in a treatment where $ \mathcal{A}_{i \Bk}(\Brho)$ is evaluated in B1  yields a similar good account of the measurements as the CDW-EIS model (except at very low  $\eta$).

\subsection{Fully differential cross sections}

A more detailed analysis can be performed on the level of the fully differential cross section (FDCS)
\begin{equation}
%\everymath{\displaystyle}
 \frac{ \mathrm{d} \sigma^{3}}{ \mathrm{d} E_k \mathrm{d}\Omega_e \mathrm{d}\Omega_{f} }  = k_e
|\mathcal{R}_{i \Bk}(\Beta)|^2,
\label{fdcs}
\end{equation}
where $\mathrm{d}\Omega_{f}$($\theta_f,\phi_f$) and $\mathrm{d}\Omega_{e}$($\theta_e,\phi_e$) denote the solid angles for the scattered projectile and emitted electron, respectively. 

Figure (\ref{fig7}) presents FDCS's for 1.5 MeV/amu O$^{8+}$ - Li(2s,2p) collisions as functions of $\phi_{e}$ when $\theta_{e}$=90$^o$, E$_{e}$=1.5 eV and q was set to 0.3 and 1.0 a.u. for 2p and 2s, respectively. The Y-Z plane is fixed by the incoming and scattered projectile's momenta with the positive Z axis pointing into the incident projectile direction. This Cartesian coordinate system is completed with an X axis to form a right-handed system. $\phi_e$ is measured in the normal way with respect to the X axis, that is Figure (\ref{fig7}) presents electron ejection cross sections in a plane perpendicular to the projectile direction. Let us first discuss results obtained with different forms of the polarization potential.

\subsubsection{Effects of polarization in the ionization of Li(2s) and Li(2p) at different q}

\begin{figure}[ht]
\centering
\epsfig{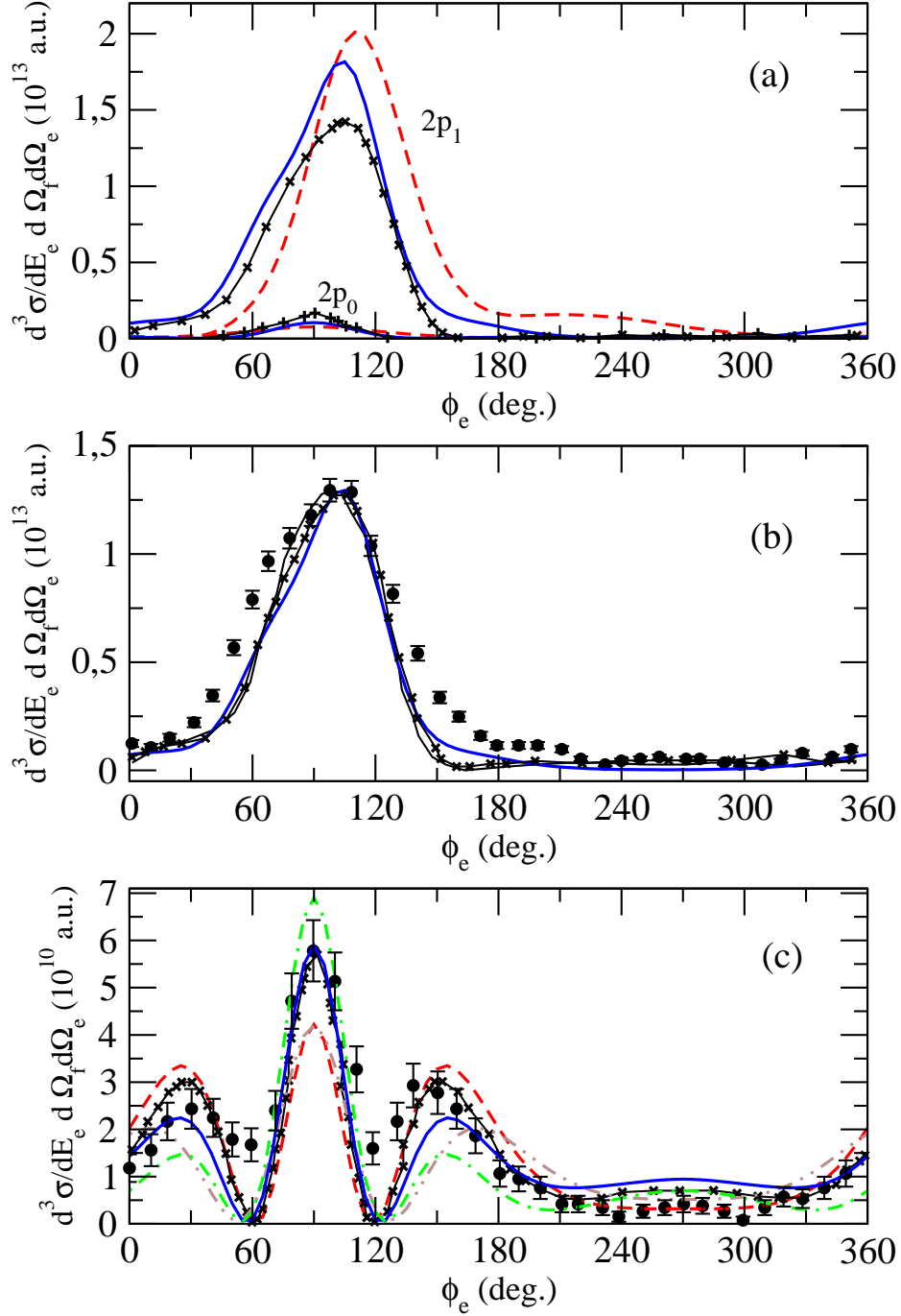}
\caption{
(Color online) Fully differential cross section  for single ionization of Li(2s,2p) by
1.5 MeV/amu O$^{8+}$ impact as a function of $\phi_{el}$ for fixed q, $\theta_{el}$=90$^o$ and E$_{el}$=1.5 eV. (a) 2p$_o$ and 2p$_1$ for q=0.3 a.u. (b) 0.3Li(2$p_o$)+0.7Li(2p$_1$) for q=0.3 a.u. (c) Li(2s) for q=1.0 a.u. Results of present calculations with NN interaction where $V_{pol}$=0 are dashed red lines and where $V_{pol}$ is given by (\ref{VP}) are thick solid blue lines; (\ref{VPExp}) are dot-dashed green lines; (\ref{VPExp-e}) are dot-dot-dashed brown lines, see the text. Thin solid lines with crosses are the CP  results from \cite{WW2014}, in (b) and (c) the CP results are multiplied by 1.2. Circles and thin black solid lines are experimental and  theoretical CDW-EIS results from \cite{Hubele2013}.
}
\label{fig7}
\end{figure}

Calculations for Li(2p) are carried out for q=0.3 a.u. and as can be expected the cross sections are not very sensitive to the form of the $V_{pol}$ interaction (only NN with (\ref{VP}) are presented in Figure (\ref{fig7}) (a) and (b)). At the same time for Li(2s) where q=1 a.u., see Figure (\ref{fig7}) (c), the FDCS is very sensitive to the form of $V_{pol}$. The calculation with (\ref{VP}) for the polarization potential, which showed a good account of the DDCS (see Figures (\ref{fig1})-(\ref{fig3})), reproduces the main characteristics of the measured distribution, however, it has defects when finer details are considered. Calculations with other forms of $V_{pol}$ are less satisfactory. Calculations  performed with an NN interaction potential using the effective target ion charge Z$_{eff}$=1.0 reveal that the collision parameters of Figure (\ref{fig7}) correspond to the distant collision regime. Similar calculations with  Z$_{eff}$=1.34 reveal better agreement in shape, but a further increase of Z$_{eff}$ cannot be justified  as it adversely affects the absolute values of the cross section. A detailed analysis shows that the relative magnitudes of the peaks in Figure (\ref{fig7}) (c) depend strongly on the character of the transition amplitude at around $\rho \approx $ 1 a.u., where the polarization potentials change drastically due to the cutting procedures.
Figure (\ref{fig7}) shows also CP results of \cite{WW2014} which, especially for Li(2s) are in better agreement with the experiment than the present calculations. Differences between the CP and CDW-EIS results appear not only in the shapes of the FDCS's but also on the absolute scale. The latter is unexpected if one recalls the good account of the DDCS by both methods, see  Figure (\ref{fig5}). We think that the observed discrepancy is related to the slightly different account of the NN interaction in the two calculations and this difference is probably emphasized with the decrease of $E_e$.

\begin{figure}[ht]
\centering
\epsfig{file=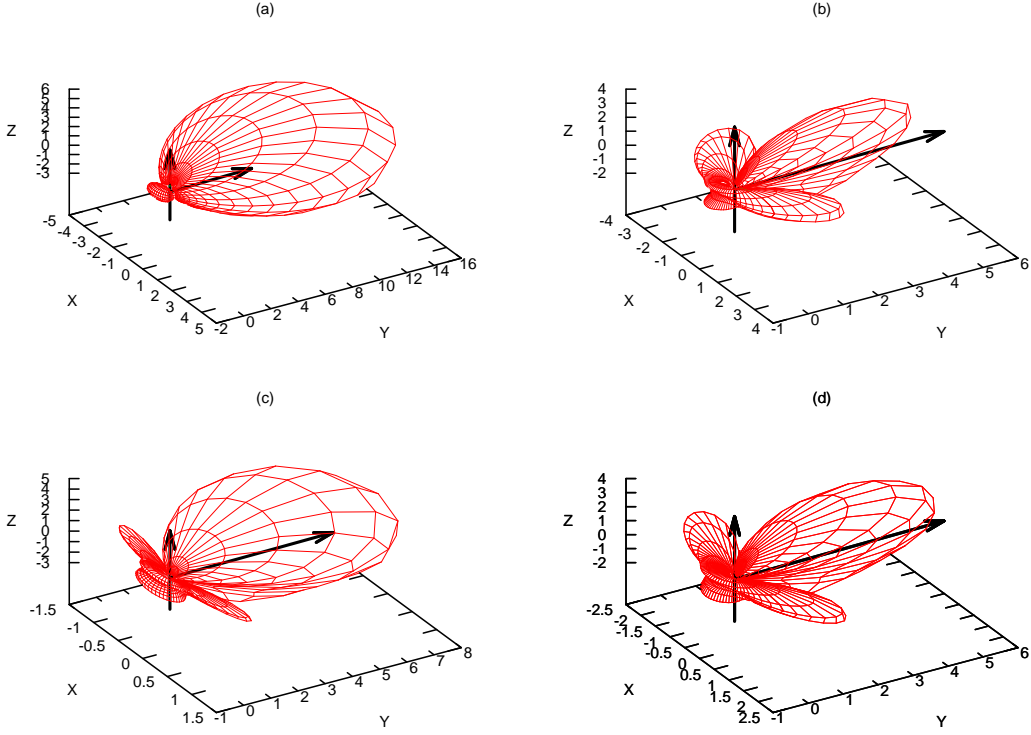,angle=-0,width=0.9\linewidth}
\caption{
(Color online) Fully differential cross section  for single ionization of Li(2s) by
1.5 MeV/amu O$^{8+}$ impact for  q=1.0 a.u. and E$_{el}$=1.5 eV.
Calculations without (a) and with NN interaction where $V_{pol}$ is given by (\ref{VP}) (b), (\ref{VPExp}) (c), (\ref{VPExp-e}) (d). The arrow pointing to toward the +Y direction denotes \textbf{q}. 
}
\label{fig8}
\end{figure}

Figure (\ref{fig8}) shows the fully differential angular distribution of electrons ejected in 1.5 MeV/amu O$^{8+}$ - Li(2s) collisions. $E_e$ is fixed at 1.5 eV and $q$=1.0 a.u. Figure (\ref{fig8}) (a) presents the FDCS evaluated without internuclear interaction, while (b)-(d) are obtained from calculations including the NN interaction with $V_{pol}$ of (\ref{VP}), (\ref{VPExp}) and (\ref{VPExp-e}), respectively. Considerable differences can be observed between results with and without NN interaction. The more characteristic differences are the sharpening of the FDCS in the direction of $\textbf{q}$ (positive Y axis) and the appearance of the wings in the $\pm$ X directions (perpendicular to the scattering plane) caused by the NN interaction. It is obvious from Figure (\ref{fig8}) that the two small peaks in Figure (\ref{fig7}) (c) at $\phi \approx$30$^o$ and 150$^o$ are due to the wings whereas calculations without NN interaction  only give rise to the centroid peak at  $\phi$=90$^o$. This has already been reported in \cite{Hubele2013} and \cite{WW2014}. At the same time considerable discrepancies are visible among results with different $V_{pol}$. The node of the 2s orbital is at around $r_T\approx$ 1 a.u.. This is the distance where the differences in the polarization potentials are emphasized. Moreover, the phase due to $V_{pol}$ affects the full NN phase only for $\rho \leq$ 1 a.u.. Accordingly,  the strong variation of the FDCS with $V_{pol}$ supports the idea that the shape of the wings is determined by the low $\rho$ character of the NN interaction.

\begin{figure}[ht]
\centering
\epsfig{file=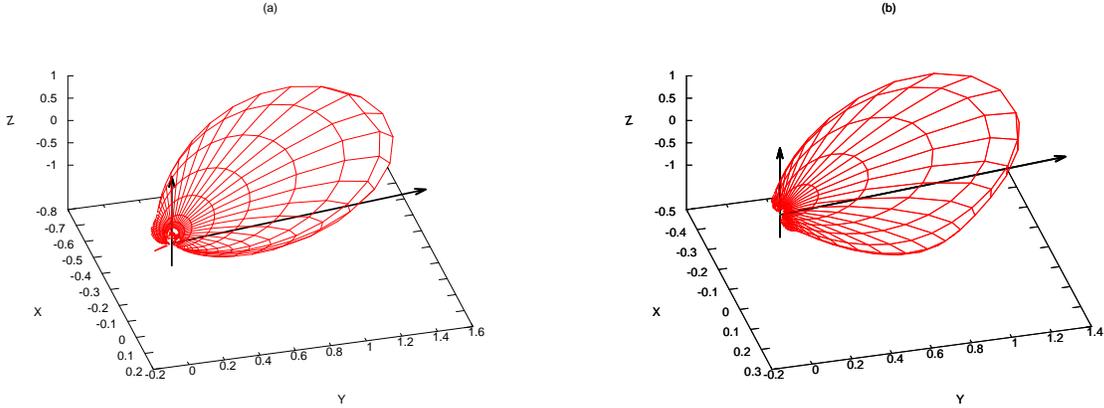,angle=-0,width=1.0\linewidth}
\caption{
(Color online) Fully differential cross section  for single ionization of 0.3Li(2p$_o$) + 0.7Li(2p$_1$) by
1.5 MeV/amu O$^{8+}$ for q=0.3, and E$_{el}$=1.5 eV.
Left panel without, right panel with NN interaction where $V_{pol}$ is given by (\ref{VP}). The arrow pointing to toward the +Y direction denotes \textbf{q}. 
}
\label{fig9}
\end{figure}

Weaker deviations appear between calculations  with and without NN interaction for the Li(2p) ionization FDCS at q=0.3 a.u., see Figure (\ref{fig9}). This FDCS is not symmetric with respect to the collision plane even for the calculation without NN interaction and the NN interaction further emphasizes this asymmetry.

\subsubsection{Satellite peaks in the ionization of Li(2s)}

\begin{figure}[ht]
\centering
\epsfig{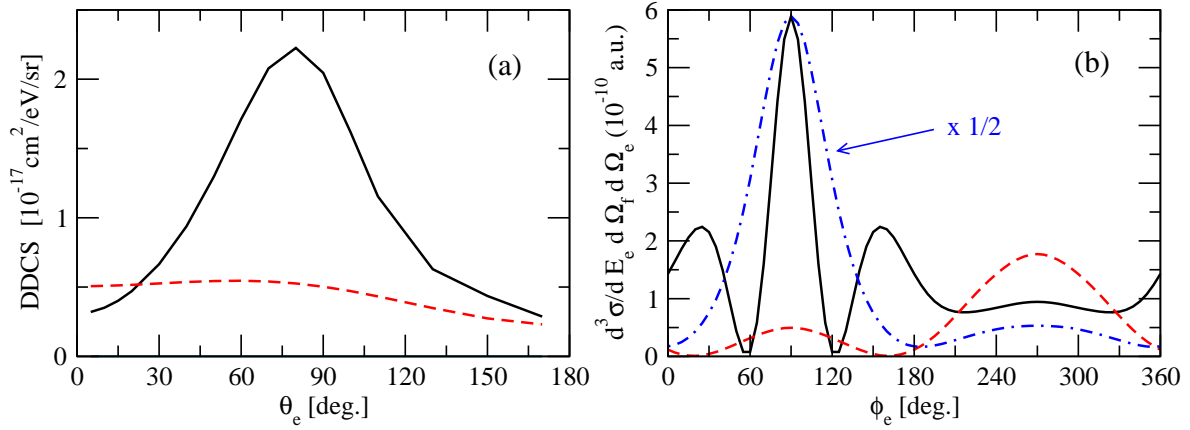}
\caption{
(Color online) Differential cross section calculated in CDW-EIS for single ionization of Li(2s) by 1.5 MeV/amu O$^{8+}$ impact for  E$_{e}$=1.5 eV. 
(a) Doubly differential cross section as a function of $\theta_{el}$. (b) Fully differential cross section as a function of $\phi_{el}$ for fixed q=1.0 a.u. and $\theta_{e}$=90$^o$. Calculations including all (solid black) and only l=0,1 (dashed red) partial waves in the expansion of the wave function of the ejected electron (see the text). In (b) the dashed-dotted blue line represents a calculation as that depicted by the solid line but without the NN interaction.
}
\label{fig10}
\end{figure}

Finally let us turn our attention to the satellite peak structure or the presence of the wings in the FDCS of Figures (\ref{fig7}) (c) and (\ref{fig8}), respectively. First of all we note that slow electrons are usually ejected in distant collisions between the projectile and target, where the three-body dipole interaction dominates \citep{sto97}. At the same time for high impact velocities and large projectile charges the two-body binary encounter mechanism plays an important role and manifests itself as a sharp peak at $\theta_e \approx$ 90$^o$ in the angular distribution. This is well seen in Figure (\ref{fig10}) (a) where the DDCS versus $\theta_e$ is presented  for $E_e$=1.5 eV. Results only due to the dipole interaction are derived from a CDW-EIS calculation where only the $l$=0,1 partial waves from the expansion of the wave function for the final state  have been taken  into account \cite{GFS95}. The definite role of the binary mechanism is well seen at $\theta_e \approx$ 90$^o$ in the Figure. The FDCS's obtained only with the dipole interaction and with all interaction terms (including dipole and binary) are displayed in Figure (\ref{fig10}) (b). This Figure shows that the multiple peak structure in the FDCS appears only when binary and NN interactions are taken into account in the calculation. The binary interaction describes a head-on collision between the electron and the projectile, which obviously is important in the region where the electron density is significant. Test calculations demonstrated that the multiple peak structure reduces to a single peak characteristic of the projectile-electron interaction, when the NN interaction is neglected in a $\rho$=[1-5] a.u. window in the calculation. This region of $\rho$ is comparable to the extension of the electron cloud for the 2s orbital. Moreover, we have performed calculations where Z$_p$ in the NN interaction has been varied. No satellite peaks are obtained for  Z$_p \leq$ 3, for which the shape of the FDCS is almost the same as in a calculation without the NN interaction. An analysis of  the classical deflection function revealed that the impact parameter that corresponds to Coulomb scattering of the projectile (with Z$_p$=8) from  the target nucleus at q=1 a.u. is at $\rho \approx$ 2.5 a.u. For $Z_p$=1 this region shifts to the much lower value $\rho \approx$ 0.2 a.u. where the electron density is negligible. These results confirm the idea that the presence of the satellite peak or the wing structure in the FDCS is due to the combination of the NN interaction and the binary collision mechanisms. This idea is further supported by the fact that a calculation performed at $q$=1 a.u. for the case of the 2p orbital shows similar satellite peak structures as discussed for 2s in Figure (\ref{fig7}) (c).

\section{Summary and conclusion}
\label{sec:conc.}

In this paper we applied the continuum distorted wave with eikonal initial state approximation to describe ionisation of Li under the impact of 6 MeV  H$^+$ and  1.5 MeV/amu O$^{8+}$ ions. Doubly  and fully differential cross sections have been evaluated within the framework of the independent electron approximation. The effect of the internuclear interaction (NN) has been taken into account  by a phase factor.     

No importance of the NN interaction has been found for the case of proton projectiles. At the same time, when describing collisions with O$^{8+}$ ions the inclusion of the NN interaction potential in the calculation cannot be avoided for a proper account of the processes at play. The NN interaction potential is made of by the Coulomb interaction of the heavy nuclei and two interaction terms due to the screening of the passive electrons and the  polarization of the target by the incident projectile ion. The most dominant effect is provided by the Coulomb interaction term, however, the inclusion of the other potentials cannot be avoided for a proper account of the processes. A characteristic role of $V_{pol}$ has been found in the DDCS for high $\eta$ values. $V_{pol}$ is not uniquely defined for short distances and accordingly different approaches and  different forms of $V_{pol}$ are available in the literature. 
Three different forms of $V_{pol}$ have been used in the present study and a very good reproduction of the measured DDCS has been obtained when $V_{pol}$ is given by (\ref{VP}). $V_{pol}$ of (\ref{VP}) depends on a cut-off parameter and its proper value might differ for the processes and systems under study. Note that we used d=0.57 for our best results, while values for d=0.3-0.93 have also been recommended in various studies in the field of electron-atom collisions \citep{Zhang1992,Mitroy2010,Yurova2014}. In the case of the FDCS deviations in different forms of the NN interaction potential are more emphasized.

The satellite peak structure observed in the FDCS for the  O$^{8+}$ - Li(2s) system were attributed to the nodal structure of the 2s orbital in \citep{Hubele2013} and \citep{WW2014}. Our study offers an alternative explanation, namely that the satellite structure is due to a combination of the NN interaction and the binary interaction mechanism.

\section{Acknowledgements}

This work was supported by the Natural Sciences and
Engineering Research Council of Canada and by the Hungarian Scientific Research Fund (OTKA Grant No. K 109440).
We thank Eberhard Engel for making his atomic structure
calculations available to us and Daniel Fischer for discussions on the data of Ref. \citep{LaForge2013}.

%We note that at high impact energies $\eta \approx 2 K\sin(\theta_P/2)$, where %$\theta_P$ and K denote,
%respectively, the scattering angle and magnitude of the momentum of the %projectile in 
%the center of mass system \cite{mcd70}.

\bibliography{/home/gulyasl/TEX/cikkek/BIB/correlation,/home/gulyasl/TEX/cikkek/BIB/Li-FDCS}

\newpage

\end{document}